\documentclass[fleqn,twoside]{article}
\usepackage{espcrc2}

\usepackage{graphicx}
\usepackage[figuresright]{rotating}
\usepackage{times}
\usepackage{graphicx}
\usepackage{epsfig}
\usepackage{xspace}
\usepackage{latexsym}
\usepackage{moreverb}
\usepackage{pifont}
\usepackage{pstcol}
\usepackage{picinpar}
\usepackage{amssymb}

\newcommand{\dee}{\mathrm{d}}

\newcommand{\AmS}{{\protect\the\textfont2
  A\kern-.1667em\lower.5ex\hbox{M}\kern-.125emS}}

\title{Loop integration results using numerical extrapolation
       for a non-scalar integral}

\author{E. de Doncker\address[WMU]{Department of Computer Science, 
        Western Michigan University, \\ 
        Kalamazoo, MI 49008, U.~S.~A.}
        \thanks{{Supported in part by National Science Foundation grants ACR-0000442, EIA-0130857,
                 ACI-0203776; Corresponding author's e-mail address: {\tt elise@cs.wmich.edu}}
                },
        Y. Shimizu\address[KEK]{High Energy Accelerator Research Organization (KEK),\\
                                Oho 1-1, Tsukuba, Ibaraki, 305, Japan},
        J. Fujimoto\addressmark[KEK],
        F. Yuasa\addressmark[KEK],
        K. Kaugars\addressmark[WMU],
        L. Cucos\addressmark[WMU],
        and
        J. Van Voorst\addressmark[WMU]
}

    \newcommand{\parint}{\textsc{ParInt}\xspace}
   
\begin{document}

\begin{abstract}
Loop integration results have been obtained using numerical integration and 
extrapolation. An extrapolation to the limit is performed with respect to a parameter
in the integrand which tends to zero. Results are given for a non-scalar four-point
diagram. Extensions to accommodate loop integration by existing integration  
packages are also discussed. 
These include: using previously generated partitions
of the domain and roundoff error guards.
\vspace{1pc}
\end{abstract}

\maketitle

{\bf Keywords.} One-loop correction, non-scalar four-point function, numerical integration,
extrapolation. 

\section{INTRODUCTION AND BACKGROUND}
In a general form, loop integrals used for cross section corrections
are given by
\begin{equation}
{\mathcal I}[\wp] = \int \prod_{\lambda=1}^L \frac{\dee^4 l_\lambda}{{(2\pi)}^4i}
\prod_{\ell=1}^N \frac{{\wp}(k_1,\ldots,k_N)}{k_\ell^2-m_\ell^2+i\varepsilon},
\label{1loop}
\end{equation}
\noindent
where $N$ is the number of propagators, $L$ the number of loops, the momentum
on the $\ell$-th internal line is $k_\ell$ and the corresponding mass is $m_\ell,$$1\le\ell\le N.$

As a special case, scalar one-loop integrals of the form $(-1)^n/(16\pi^2)I_n$ where
\begin{equation}
$$
I_n = \int_{{\mathcal S}_{n-1}} \frac{1}{(D_n({\bf x})-i\varepsilon)^{n-2}}
\dee {\bf x}
$$
\label{1looppar}
\end{equation}
are obtained from~(\ref{1loop})
by introducing Feynman parameters
and integrating over the loop momentum $l$.
The integration region $S_{n-1}$ is the $n-1$ dimensional unit simplex.
 
For the simplest cases, the results can be obtained analytically.
Numerical techniques have been successful with 
considerable analytic manipulation (see,
e.g.,~\cite{numerical,ferroglia03}). In previous work~\cite{edcpp03},
we reported results for integrals of the form~(\ref{1looppar}) treated 
numerically using extrapolation by the $\varepsilon$-algorithm~\cite{wynn56}.
We will now consider the case of a one-loop integral where the numerator in the integrand is a
polynomial of the Feynman parameters. A sample problem involving the
$e^- e^+ \rightarrow W^- ~W^+$ interaction
is given in the next section of this paper. Results for this problem 
are given in Section~\ref{results}.
Section~\ref{parint} discusses enhancements to the ParInt parallel
integration package.

\section{NON-SCALAR INTEGRAL}
\label{non-scalar}

\begin{figure}
\begin{center}
\includegraphics[clip,width=1.1\linewidth]{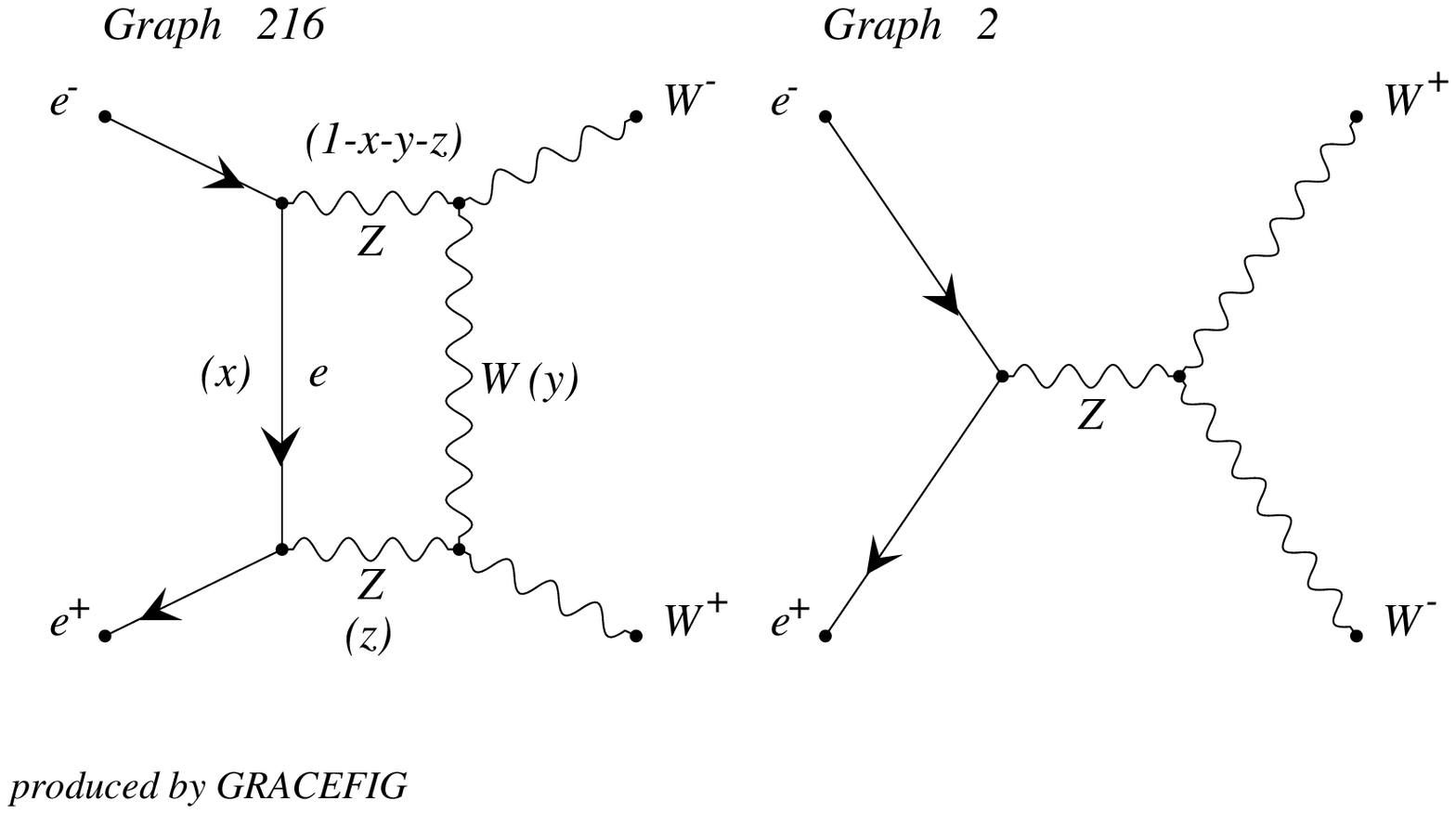}
\caption{Feynman diagram for $e^- e^+ \rightarrow W^- ~W^+$}
\end{center}
\label{gr216}
\end{figure}

The matrix element of one-loop corrections is given by the real part of the
product of a one-loop amplitude and the (conjugate of) a tree amplitude.
Figure~\ref{gr216} shows an example of a box diagram and a tree diagram of
a $Z$-boson
exchange for the interaction $e^- e^+ \rightarrow W^- ~W^+.$
The Feynman diagram and the corresponding matrix element are generated
automatically by {\tt GRACE-loop}\cite{graceloop} system. 

After introducing the
Feynman parameters as in Figure~\ref{gr216}, and integrating over the
loop momentum,
the matrix element is of the following form,
$$
M_4(f,g;\varepsilon) = \int dx dy dz \left[  \frac{f(x,y,z)}{(D_4-i\varepsilon)^2}
                     - 2\frac{g(x,y,z)}{D_4-i\varepsilon} \right]
$$
where
$D_4 = {~}^\tau {\bf x} A {\bf x}+2{\bf v}\cdot {\bf x}+C,$
and $A_{\iota j} = q_\iota\cdot q_j, ~q_1 = -p_{e^-},
~q_2 = p_{e^+}, ~q_3 = p_{e^+}-p_{W^+}$,
$C = M_0^2 = M_Z^2$,
$v_\iota = \frac{1}{2}(-q_\iota^2 + M_\iota^2 - M_0^2)$ with
$M_1 = m_e,M_2 = M_W,M_3 = m_e$.
                                                                                        
\begin{figure}[ht]
\begin{center}
\fbox{\rotatebox{-90}{
\includegraphics[clip,width=.65\linewidth]{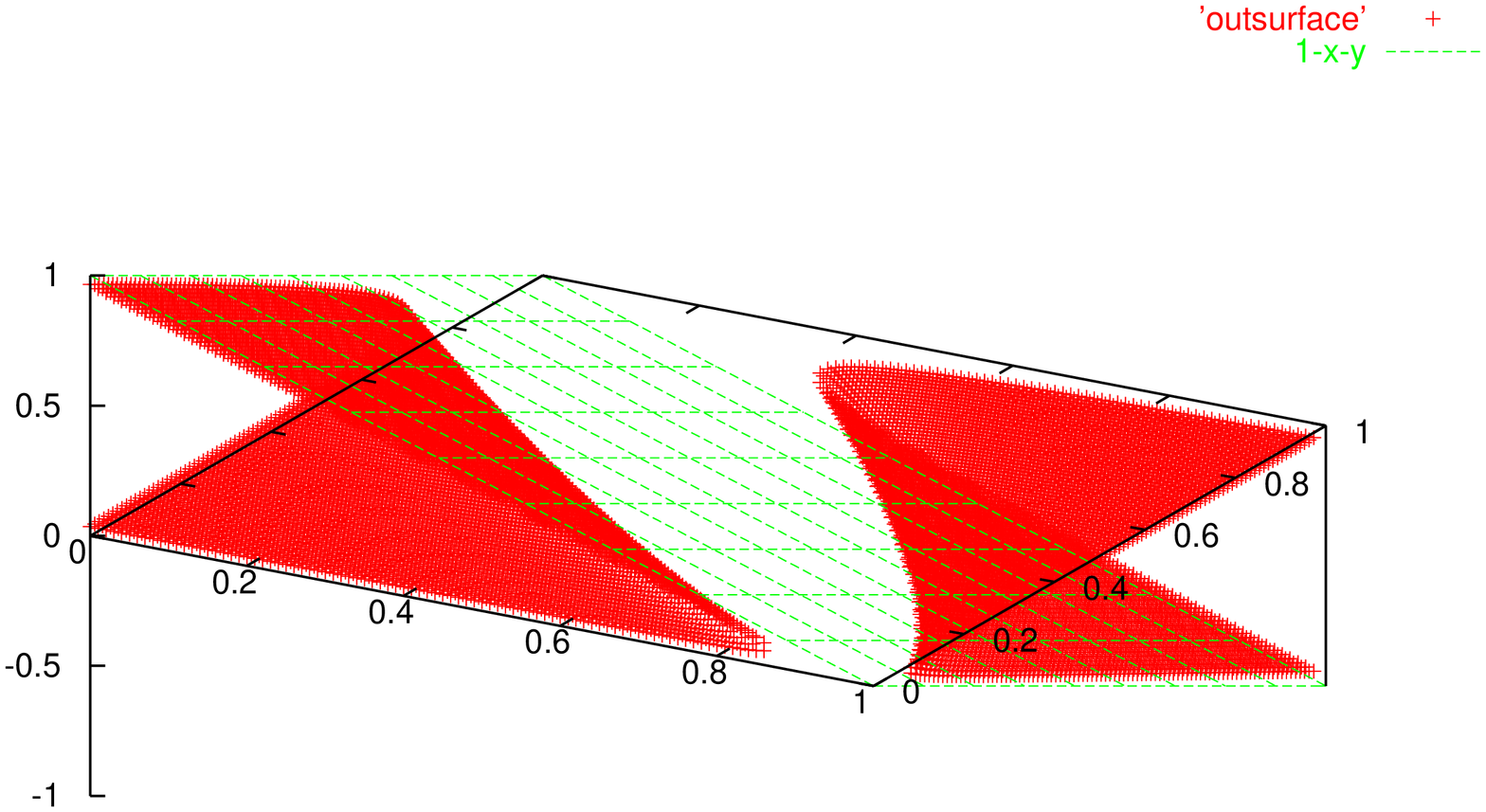}
}}
\caption{$D_4 = 0$ surface
}
\label{x4.ps}
\end{center}
\end{figure}

Figure~\ref{x4.ps} shows the $D_4 = 0$ surface of the singularity
over $-1 \le x, y \le 1,$ and delineates
the integration domain ${\mathcal S}_3.$

%
$f$ and $g$ are polynomials of Feynman parameters, of which the coefficients
are determined by external momenta and masses of internal lines.
Here $M_Z = 91.187 {\rm GeV}, M_W=80.22 {\rm GeV}, m_e = 0.511 {\rm MeV},
~\sqrt{s} = 500~ {\rm GeV}$ ~and~
$\theta = \angle ({\bf p}_{e^-},{\bf p}_{W^-})$. 
The numerical results are evaluated for $\cos\theta = 0.956811390.$ 

The generalized
non-linear gauges\cite{graceloop} are implemented for the amplitude.
The result depends on the gauge parameters because only one diagram is
picked up. For the numerical evaluation, the non-linear gauge parameters
are set as $\tilde{\alpha}=2,\tilde{\beta}=3, \tilde{\delta}=4,
\tilde{\epsilon}=5$ and $\tilde{\kappa}=6$.

\section{GRAPH 216 RESULTS}
\label{results}

Table~\ref{res216x2} illustrates the use of the $\varepsilon$-algorithm
for the integral computation of the term involving $f$ (the symbolic code
of which has about 2000 lines as FORTRAN code).
We show the results of the extrapolation for the real part of $M_4(f,0;\varepsilon).$
The method is based on generating a sequence of integral values corresponding
to a geometric sequence of $\varepsilon$ and extrapolating to the limit as 
$\varepsilon \rightarrow 0.$

The table shows the sequence of integral approximations for $\varepsilon = 1.2^{30-\ell},~~
\ell = 0,1,\ldots$ (obtained numerically) in the first (leftmost) column. Using the
integral approximations corresponding to $\ell = 0, 1, 2,$ the first extrapolated result is obtained
(top element of column 2). Using the $\ell = 3$ element of column 1, the second extrapolated
result is obtained in column 2; the $\ell = 4$ element of column 1 is then used to 
generate the third element of column 2 and the top element of column 3. In all iterations
following, the new element in column 1 is used to generate a new lower diagonal of the 
triangular table. 

The table elements are shown to 8-digit accuracy, which is the final accuracy obtained
in this run. Convergence is apparent down the columns and along the lower diagonal. Relying
on a heuristic error estimate of the table elements along the lower diagonal, an element is selected as the result (printed boldface). The result calculated analytically is -0.647837287.

\begin{sidewaystable}
\caption{Extrapolation table for the real part of $M_4(f,0;\varepsilon)$}
\newcommand{\m}{\hphantom{$-$}}
\newcommand{\cc}[1]{\multicolumn{1}{c}{#1}}
{\footnotesize
\begin{tabular*}{\textheight}{@{\extracolsep{\fill}}lllllllllllll}
\hline
& $\ell$  & & & & & & & & & &   \\
\hline
& 0 & -0.65001198 & & & & & & & &  \\[2pt]
& 1 & -0.65102708 & -0.65168630 & & & & & & &  \\[2pt]
& 2 & -0.65142675 & -0.65143432 & -0.64911741 & & & & & &  \\[2pt]
& 3 & -0.65143418 & -0.65142699 & -0.64873688 & -0.64669569 & & & & &  \\[5pt]
& 4 & -0.65121543 & -0.65186148 & -0.64837272 & -0.64766170 & -0.64787059 & & & &  \\[5pt]
& 5 & -0.65088469 & -0.65400251 & -0.64811405 & -0.64780004 & -0.64784821 & -0.64784032 & & &  \\[2pt]
& 6 & -0.65051471 & -0.61144371 & -0.64796871 & -0.64783212 & -0.64784301 & -0.64784011 & -0.64784042 & &  \\[2pt]
& 7 & -0.65014820 & -0.64536958 & -0.64789843 & -0.64783979 & -0.64784142 & -0.64783917 & -0.64783494 & -0.64783735 &  \\[2pt]
& 8 & -0.64980779 & -0.64694643 & -0.64786686 & -0.64784110 & -0.64784083 & -0.64783800 & -0.64783707 & -0.64783723 & -0.647837056 \\
& 9 & -0.64950358 & -0.64742949 & -0.64785289 & -0.64784075 & -0.64784147 & -0.64783737 & -0.64783720 & {\bf -0.64783719}  \\
& 10& -0.64923827 & -0.64763299 & -0.64784650 & -0.64784000 & -0.64782758 & -0.64783723 & -0.64783718 &   \\
& 11& -0.64901060 & -0.64773083 & -0.64784332 & -0.64783920 & -0.64783650 & -0.64783719 &  &   \\
& 12& -0.64881731 & -0.64778110 & -0.64784155 & -0.64783848 & -0.64783702 &  &  &   \\
& 13& -0.64865441 & -0.64780784 & -0.64784045 & -0.64783791 &  &  &  &   \\
& 14& -0.64851780 & -0.64782228 & -0.64783970 &  &  &  &  &   \\
& 15& -0.64840361 & -0.64783009 &  &  &  &  &  &   \\
& 16& -0.64830839 &  &  &  &  &  &  &   \\
\hline
\end{tabular*}\\[2pt]
}
\label{res216x2}
\end{sidewaystable}

To generate the integral approximation in the first column, we used an iterated
integration where the adaptive Quadpack~\cite{pi83} routine {\sc dqage} was used in each 
direction, requesting a relative accuracy of $10^{-10}.$ So far, this technique
has outperformed other numerical integration approaches using multivariate (cubature)
rules. 

\section{PARINT ENHANCEMENTS}
\label{parint}
ParInt is a software package for parallel multivariate integration~\cite{parintweb}.
It has components for multivariate integration using Monte Carlo ({\sc mc}), Quasi-Monte Carlo
({\sc qmc}) and adaptive methods. ParInt is written in C and runs over MPI~\cite{MPIweb}
on a distributed platform. 

\subsection{Iterated integration}
While the adaptive approach could not be applied directly using 3D multivariate rules, 
results to 6-figure accuracy were obtained by treating the problem as a 2D integration
of a 1D integral. The 1D inner integral was calculated with Quadpack routine {\sc dqage}.

\begin{figure}
\begin{center}
\fbox{\includegraphics[clip,width=.9\linewidth]{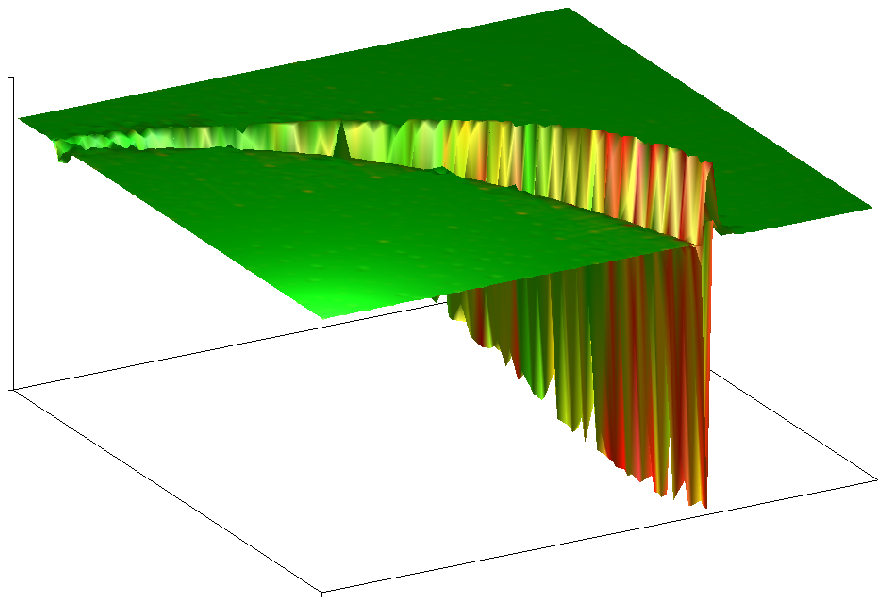}}
\caption{2D of 1D integrand for Graph 216 real part
}
\label{drawcq.ps}
\end{center}
\end{figure}

The 2D integration was performed with ParInt and with its Fortran sequential predecessor,
{\sc dcuhre}~\cite{gnzbe91a}. The local region error estimate was changed to make it 
less conservative.
Figure~\ref{drawcq.ps} illustrates the integrand of the
2D problem for $\varepsilon = 1.2^{25},$ which was drawn using evaluation points of the
integration.
We are currently considering a design of ParInt which will allow incorporating iterated integration
in a transparent way.

It recently came to our attention that in the work by Binoth, Heinrich and Kauer~\cite{binoth02},
3D box integrals are obtained by performing the inner integration analytically, which
leaves the resulting 2D integrand with an integrable (though still problematic)
singularity. Note further that their 3D box together with 2D vertex diagram evaluations 
are at the basis of reductions performed to treat scalar hexagon integrals.

\subsection{Re-use of subregions between extrapolations}
A sequence of extrapolation steps uses a series of similar integrations
which share similar subregions when performed adaptively.  At each step of the
extrapolation, \parint can avoid a significant number of region
evaluations by re-using previously ``discovered'' subregions as the
initial set of regions for the next extrapolation step, potentially
avoiding a significant amount of computation.

\parint has been modified to support this activity by storing the
active integration regions at the end of every parallel run and
providing the user with the option to load a set of regions to
initialize a run.  Regions may be saved either locally on each compute
node or in a single file managed by the controller.  Regions loaded at
the start of a subsequent computation may also be read from a single
global file or individual files on each compute node.  We are
currently developing a distributed I/O system which will allow compute
nodes to retrieve previously saved regions from files on any other
compute node~\cite{lcucos03}.

\subsection{Kahan summation}
The global adaptive integration algorithm first developed by 
De Ridder and Van Dooren~\cite{deridder76} is also used by ParInt.  At
each step, one region (per worker) is selected and subdivided into
subregions.  The selected integration rule is applied over each
subregion.  Next, the estimated error and result for the selected
region and subregions must be subtracted from and added to the total
estimated error and result, respectively.  For difficult problems,
ParInt will select many regions and subdivide them.  Numerical
summation of millions of terms can introduce round off error and
greatly reduce the accuracy of the result and estimated error in a
numerical integration routine.

We have looked at several techniques to reduce round off error in 
sums with a large number of terms.  Each of these techniques has its
own merits and flaws.  A good method would be one whose accuracy does 
not depend on the number of terms in the sum and would not greatly
impact the runtime performance of a numerical integration routine.
A compensated summation method developed by W. Kahan~\cite{kahan65} 
and further studied by N. Higham~\cite{higham93}
best fits these needs.  Several advantages of this method are
low computational overhead, low storage requirements, and in error
analysis it is shown to have an error constant of order 1.

\section{CONCLUSIONS}
We presented results for a non-scalar one-loop box diagram, where the 
integral is obtained using numerical integration and extrapolation
with the $\varepsilon$-algorithm. 
We described enhancements to the ParInt parallel integration package,
which are in various stages of development.
Furthermore, in future work, we plan to investigate combinations of
our numerical methods with symbolic techniques.

\end{document}